# Dielectric and electro-mechanic nonlinearities in perovskite oxide ferroelectrics, relaxors and relaxor ferroelectrics

Lukas M. Riemer[1,a)], Li Jin[2], Hana Uršič[3], Mojca Otoničar[3], Tadej Rojac[3], and Dragan Damjanovic[1].

a)Author to whom correspondence should be addressed: lukas.riemer@epfl.ch

[1]Group for Ferroelectrics and Functional Oxides, Swiss Federal Institute of Technology in Lausanne – EPFL, 1015 Lausanne, Switzerland
[2]Electronic Materials Research Laboratory, Key Laboratory of the Ministry of Education & International Center for Dielectric Research, School of Electronic and Information Engineering, Xi'an Jiaotong University, Xi'an 710049, China
[3]Electronic Ceramics Department, Jožef Stefan Institute, Jamova Cesta 39, 1000, Ljubljana, Slovenia

**Abstract**
The polarization and strain response of ferroelectric materials at fields below the macroscopic coercive field is of a paramount importance for the operation of many electronic devices. The response of real ferroelectric and related materials is in general complex and difficult to interpret. The reason for this is that many processes in a ferroelectric material contribute to its properties, often concurrently. Examples include motion of ferroelectric and ferroelastic domains, presence of domains within domains, dynamics of different types of polar nano-entities, interaction of polar nano-entities (e.g., polar nanoregions in relaxors) with the strain and polarization within domains, motion of defects and rearrangement of defect clusters and their interaction with polarization and strain. One signature of these processes is nonlinearity of the strain and polarization. Most ferroelectrics exhibit nonlinear response at all practical field levels meaning that the apparent material coefficients depend on the amplitude of the driving excitation. In this paper we show that an investigation of nonlinear behavior is a sensitive way to study various mechanisms operating in dielectric and piezoelectric materials. We review the basic formalism of the nonlinear description of polarization and strain, give a physical interpretation of different terms and illustrate this approach on numerous examples of relaxors, relaxor ferroelectrics, hard and soft ferroelectrics, and morphotropic phase boundary compositions. An experimental approach based on a lock-in technique that is well-suited for such studies is also discussed.

## I. INTRODUCTION

Modern technology demands improved functionality, reliability and tolerances of electronic components. Dielectric and piezoelectric nonlinearities present opportunities and challenges in this regard. From the application point of view, nonlinearities can be desirable, for example in tunable filters and antennas, positive temperature coefficient resistors, variconds, non-linear optics, and ferroelectric memory; or undesirable, as in precision actuators and sensors, high frequency transducers, accelerometers and capacitors. An enhanced response is often connected to nonlinearities and reduced temporal stability. Especially in the field of ferroelectrics, where materials are tailored to take advantage of structural instabilities, non-linearities can arise on multiple time and length scales. The fundamental understanding of underlying mechanisms and their interactions can be exploited to implement innovative functionalities and to reveal active mechanisms in novel materials.

In dielectric materials that can be sufficiently well described by intrinsic lattice response (e.g., movement of atoms around equilibrium positions) rather weak nonlinearities are observed at not too large fields.[1] For most practical purposes such materials are considered as linear. Examples are sapphire or strontium titanate at room temperature, and piezoelectric materials like quartz or aluminum nitride. However, when defects, dipolar clusters, domains, or other features in a material can readily respond to the field, their response is often nonlinear. With appropriate analysis, these nonlinearities may become visible even at weak fields where the linear lattice response is dominant. Rather strong nonlinearities may appear in ferroelectrics, relaxors, or relaxor ferroelectric solid solutions and are correlated to the motion of ferroic domain walls and mesoscopic polar structures.[2] Nonlinearities in dielectrics and ferroelectrics have been studied intensively[3,4,5] while the interest in nonlinear behavior of relaxor and relaxor-ferroelectrics is more recent.[6,7,8,9] By studying nonlinearities in materials' properties it is possible to learn about the microscopic objects, their dynamics and processes that cause them.

In this work, we discuss a large spectrum of different experimentally observed nonlinear behaviors in ferroelectric and relaxor-based materials and give, when possible, physical interpretation of the nonlinear data in terms of microscopic mechanisms. A simple formalism is presented, which permits to draw some conclusions on dynamic processes taking place in the material even when underlying physical mechanisms are not well understood.

## II. NONLINEAR DESCRIPTION

One of the most fundamental periodic time-dependent signals and a common excitation signal for physical experiments is a single sinusoid:

$$F(t) = F_0 \sin(\omega t), \tag{1}$$

where $F(t)$ is the excitation signal (a driving field), $t$ is time, $F_0$ is the excitation amplitude, and $\omega$ is the angular frequency. In this paper, $F$ is the electric field, $E$. A linear system will respond to this type of excitation with a sinusoidal response of the same frequency and may be shifted in time by a phase angle $\delta$ with respect to the field:

$$R(t) = R_0 \sin(\omega t) = m F_0 \sin(\omega t + \delta), \tag{2}$$

where $R(t)$ is the response, $R_0$ is the response amplitude and $m$ is a system characteristic coefficient that quantifies the linear response and which may be frequency dependent.[10] For a linear system, $m$ and $\delta$ are independent on the driving field amplitude. For the present purposes, $R$ is electric polarization or strain, and $m$ dielectric permittivity or piezoelectric coefficient, respectively. In general, however, the relationship between the field and response is more complex than (2) and $m$ and $\delta$ can be field dependent. The material is then said to be nonlinear with respect to the field. Often, $R(F)$ is both hysteretic and nonlinear.[11]

In principle, the $R(F)$ function contains all field-dependent information on the system and can be conveniently displayed on an oscilloscope, either in form of two separate time dependent functions, $R(t)$ and $F(t)$, or in form of so-called hysteresis loop, $R(F)$. Polarization– and strain–electric field relations, which are of interest here, can come in many forms and shapes.[12] Hysteretic relations present multiple values of $R$ for most electric field levels. The nonlinear, arbitrary and nonanalytical nature of $R(F)$ loops makes analysis, comparison and quantitative evaluation of data cumbersome. An approach to overcome these limitations is to analyze data in the frequency domain, where analytical functions make the quantitative evaluation of data more convenient.

Fourier's theorem states that arbitrary periodic functions can be represented as series of weighted sinusoids and cosinusoids that are equidistantly spaced in frequency domain by $\Delta\omega = 2\pi/T$, where $T$ is the period of the arbitrary function.[13] The sinusoid with angular frequency $\omega = 2\pi/T$ is called the fundamental or first harmonic component. Waveforms with angular frequency $\omega_n = n\omega$ are called the $n^{th}$ harmonic component. For electric driving fields, $E(t)=E_0\sin(\omega t)$, a general response function $R(t)$, where $R(t)$ stands for polarization, $P(t)$, or strain, $x(t)$, can be derived in form of a Fourier series[14,15]:

$$R(t) = R_{const} + \sum_{n=1}^{\infty} R^{(n)} \sin(n\omega t + \delta_n)$$
$$= R_{const} + \sum_{n=1}^{\infty} R'^{(n)} \sin(n\omega) + R''^{(n)} \cos(n\omega)] \qquad (3)$$

with:

$$R^{(n)} = \sqrt{(R'^{(n)})^2 + (R''^{(n)})^2}, \qquad (4)$$

$tan\delta_n = \frac{R''^{(n)}}{R'^{(n)}}, R''^{(n)} = R^{(n)}\cos(\delta_n), R'^{(n)} = R^{(n)}\sin(\delta_n),$ \qquad (5)

and

$m'^{(n)} = \frac{R'^{(n)}}{E_0}, m''^{(n)} = \frac{R''^{(n)}}{E_0},$ \qquad (6)

where $R_{const}$ is a constant (a "dc" term), $R^{(n)}{}_n$ is the amplitude of the $n^{th}$ harmonic, $\delta_n$ is the phase angle, $R'^{(n)}$ the amplitude of the in-phase component or real part, $R''^{(n)}$ the amplitude of the out-of-phase (quadrature) component or imaginary part of the response, $m'^{(n)}$ the real part, and $m''^{(n)}$ the imaginary part (or loss) of the material coefficient. In general, $m$ is known as the material's susceptibility (here, dielectric susceptibility or piezoelectric coefficient). Note that the sign of the phase angle is determined by the choice of the sign in the argument of the sinusoidal function in (3); in our case it is a plus sign. Note that $R'^{(n)}$ and $R''^{(n)}$ (and thus also $\delta_n$) are, in general, functions of $E_0$.[5] As is usually the case, we define, the dielectric permittivity, $\varepsilon^{(n)}$, and piezoelectric coefficient, $d^{(n)}$, as:

$$\varepsilon^{(n)} = \frac{P^{(n)}}{E_0}, \tag{7}$$

and

$$d^{(n)} = \frac{x^{(n)}}{E_0}, \tag{8}$$

where $P^{(n)}$ and $x^{(n)}$ are amplitudes of the $n^{th}$ polarization and strain harmonic, respectively, and $E_0$ the driving field amplitude. Note that a simplified notation that emphasizes only the order of the harmonic is used and matrix or tensor notation are omitted.[16] The last two would be redundant in this case since only the longitudinal measurement mode, were polarization and strain are measured in the direction of the electric field, is discussed throughout the manuscript.

Only noncentrosymmetric materials may be piezoelectric, therefore the first harmonic and all odd number higher harmonics should not appear in strain in centrosymmetric materials (for example, unpoled ceramics). For the present purposes we shall not discuss separately electrostrictive the effect,[17] $x(E) = ME^2$, where $M$ is the electrostrictive coefficient, which appears in all dielectric materials and will consider it only within the total strain response, together with the piezoelectric effect. Similarly, it is usually assumed that in centrosymmetric materials the induced polarization should not exhibit even number harmonics.[10,18,19] However, materials of interest here are rarely ideal and both odd and even harmonics in general appear in the Fourier expansion of both polarization and strain responses. We will come back to this point again in the subsequent sections.

In principle it is possible to measure the time-dependent response of a system, $R(t)$, and mathematically perform a Fourier transformation to obtain harmonic components. An elegant experimental alternative is the use of lock-in techniques. Lock-in techniques are capable of filtering small signals of a specific frequency out of large background noise. Tunable band pass filters with quality factors above $10^8$ and a noise bandwidth below 0.001 Hz can be achieved by taking advantage of the orthogonality of sine and cosine functions.[20] This makes lock-in amplifiers effective tools to directly extract harmonic components.

This manuscript presents a detailed experimental and phenomenological description of the harmonic analysis of dielectric and piezoelectric nonlinearities in selected materials, its current understanding, and challenges. The manuscript, which presents and discusses previously unpublished data, is written with an extended introduction to aid researchers, who may not be familiar with the nonlinear analysis of dielectric and piezoelectric properties. A simple example of data acquisition software is provided in the supplementary material.

A word of caution may be necessary at this point. Electronic instruments (signal generators and amplifiers) can add distortions and a parasitic direct voltage ("dc") bias to driving signals. A distorted ac driving signal necessarily adds higher harmonics to the effective material output signal, even for perfectly linear materials. The dc bias generates a first harmonic (piezoelectric-like) strain response even if the material is centrosymmetric in the nominally unbiased condition.[21] While instrumental distortions may appear to be small when looking at the instrument specifications, they can be significant when investigating thin films, where small signals translate into large fields, and in bulk materials, which often require amplification of the signal. Also, amplifiers will act as filters and may attenuate dynamic signals at higher frequencies, distorting both phase and amplitude of higher harmonics.

**III. HARMONIC ANALYSIS**

As mentioned previously, lock-in techniques are capable of directly measuring fundamental and harmonic components of signals. When applied to polarization or strain measurements they can thus be used to obtain Fourier expansions of polarization- or strain-electric field hysteresis loops. A single hysteresis loop in the polarization or strain versus electric field plane is thereby represented by a set of amplitudes and phase angles (Figure 1). In this representation the shape of hysteresis loops can be fully quantified if a sufficient number of harmonics are used. For the basic analysis of hysteresis loops at sub-coercive driving fields the first three harmonics (n =1,2,3) are usually sufficient. More complex hysteresis loops however may require additional terms.[22]

Being able to determine and track slight changes in the shape of hysteresis loops to varying excitation levels enables detailed studies of electric field-dependent nonlinearities and their underlying mechanisms (such as motion of domain walls and polar nano-entities). To accomplish this, a phenomenological model is needed, which makes definitive predictions of nonlinear behavior. The model can then be verified experimentally by measuring harmonic components. The best-known example is the Rayleigh-Néel[23,24] model for domain wall motion in a random potential, which will be discussed in some details in the next sections. In that case only odd harmonics are present in Fourier expansion and the phase angle of all harmonics is -90°. Therefore, the Fourier analysis allows for a relatively straightforward verification of this model. However, even if a model is not available, which is most often the case especially when the nature of mesoscopic objects is not well understood, at least some details of the dynamic behavior of the hypothetical mesoscopic objects can be deduced from the nonlinear experimental data. Such examples will be discussed in the subsequent sections for relaxors and acceptor doped ("hard") ferroelectrics.

The results of Fourier analysis can be interpreted in three ways. The first is based on the geometric description of harmonic components, which is always possible. The second is the interpretation based on existing models and theories that apply for that particular system. The theory predicts a definite nonlinear response which can then be verified experimentally. While such theoretical background exists for some types of motion of ferroic domains (for example motion of domain walls in soft ferroelectrics), it is missing or is insufficiently developed for other polar objects (polar nanoregions, defect clusters) and their interactions with ferroic domains. The third way is a phenomenological interpretation, where materials under investigation and their response are compared to well-understood systems.

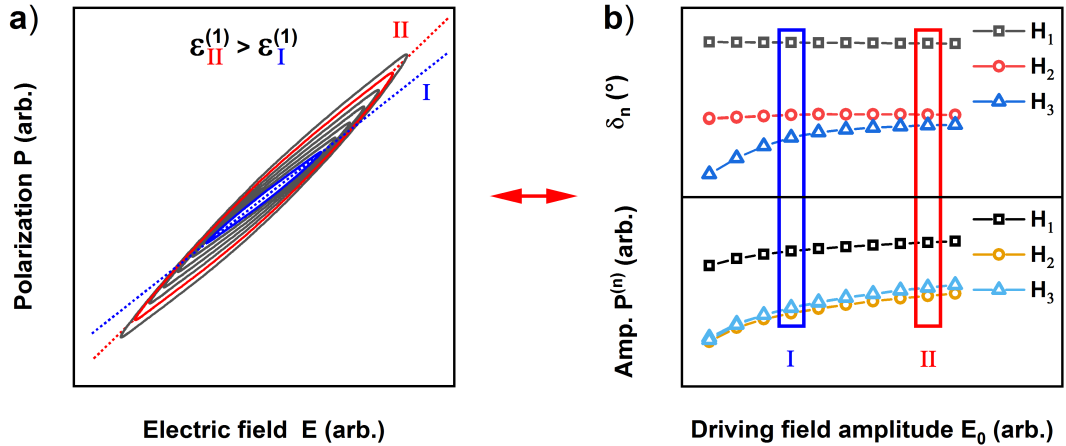

Figure 1. P(E) hysteresis loops and Fourier expansion of $P(t)$ measured in a PZT ($Pb(Zr_{1-x},Ti_x)O_3$) sample : a) Polarization hysteresis loops in polarization vs. electric field space for different maximum electric field amplitudes $E_0$. b) Amplitudes and the phase angles of the first three harmonics ($H_n$) of the corresponding Fourier expansion of the hysteresis loops shown in a). The components of the Fourier series are determined experimentally using a lock-in technique. The red and blue hysteresis loops in a) are represented by the set of coefficients framed in red and blue in b). Only the first three out of an infinite number of harmonics are given. Dotted lines in a) illustrate the linear response which is contained in the amplitude of the first harmonic $P^{(n)}$ (see discussion on Rayleigh law below and equation (9)).

**IV. GEOMETRIC INTERPRETATION**

To understand how different harmonics affect response of a material, we start with a purely formal, geometric approach, which may still reveal important details of nonlinear dynamic processes. We strongly recommend to the reader to familiarize with this approach by individually plotting fundamental (n=1) and harmonic components (n>1) as a function of a sinusoidal driving field as shown in Figure S1.

The resulting curves represent Lissajous curves with frequency ratios 1:n, where n is the order of the harmonic. In general, odd harmonics represent antisymmetric signals and even harmonics symmetric response. Due to the orthogonality of sinusoids and cosinusoids only the out-of-phase component of the first harmonic contributes to the area of the hysteresis, whose physical meaning may be related to energy loss.[11,25,26] All other harmonics merely change the shape of the overall hysteresis loops. In particular, in-phase component of odd harmonics ($\delta_n=0°$) and components with $\delta_n =180°$ for n>1 bend hysteresis ~~loops~~ (most notably the tips of the loops) while out-of-phase components redistribute the hysteresis area. For even harmonics the opposite is true. Out-of-phase components ($\delta_n =90°$ and -90°) bend hysteresis loops while in-phase components redistribute hysteresis area. The obtained curves have special symmetry relations that depend on the parity of the harmonic component.

It is seen that the corresponding Lissajous curves of odd harmonics are antisymmetric with respect to the abscissa, while the Lissajous curves for even harmonics are symmetric with respect to the ordinate. The only difference is the sense of rotation of these loops, defining which portion of the total hysteresis loop is expanded and which part is contracted by a given harmonic. As will be seen later, this is a defining feature of so-called "hard" (acceptor-doped) perovskite oxide ferroelectrics.[22] Even though it is mathematically possible to obtain a clockwise rotation of the (first harmonic) strain- and polarization- electric field hysteresis loops (for phase angle 0°< $\delta_n$ <180°) it is thermodynamically forbidden for polarization, as it would represent a net gain of energy.[26] This limitation does not apply for electro-mechanical loops, with mixed mechanical and electrical functions, such as $x(E)$, where the area of the hysteresis does not have energy units[26]

The first harmonic or fundamental component contains the linear response of a system (and possesses, in general, nonlinear terms as well). In the case of the data presented in Figure 1, $\varepsilon^{(1)}$ can be calculated following Equation 7 as the quotient of $P^{(1)}$ and $E_0$ while $d^{(1)}$ could be calculated from the corresponding strain measurements. It is seen that the dielectric permittivity increases with increasing electric field amplitude of the bipolar cycle, as one would expect for a nonlinear material. Mechanisms contributing to hysteresis can be of different origins, including conductivity, displacement of ferroelectric and ferroelectric-ferroelastic domain walls and dipole reorientation. In the case of the data in Figure 1, the hysteresis loops are dominated by irreversible domain wall motion.[27] The phase angle of the first harmonic reflects the symmetry of the corresponding tensor property. In this sense the phase angle of the first harmonic in polarization for a dielectric without loss has to be zero, while the phase angle of the first harmonic in strain for a material without loss can be either 0° or 180°. The phase angle of the first harmonic in strain depends on the crystallographic orientation and, in the case of ferroelectric materials, on the poling direction.

The second harmonic represents response that is invariant to the sign of the applied excitation field. Switching of 180° domains and other polar dipoles can contribute to second harmonics response.[17] In that case, the amplitude of the second harmonic in strain is directly related to the electrostrictive response while the second harmonic in polarization leads to asymmetry of polarization. Switching of 180°domains and piezoelectric effect together may lead to the so-called "butterfly" loops in strain which are sometimes interpreted as electrostrictive effect.[28] Incidentally, it has been shown that displacement of 180° domain walls may contribute to the piezoelectric effect if polarizations on two sides of the domain wall are not equal.[8] This is a reasonable assumption in thin films and polycrystalline materials due to complex internal field distribution.

As the second harmonic in strain represents electrostriction, it is always present regardless of the symmetry of the material. The ideal case of pure intrinsic electrostriction results in a second harmonic phase angle of -90° for positive or +90° for negative[29] electrostriction coefficients independent of the sample orientation or poling direction. It can be shown that in ideally centrosymmetric materials the second and all higher even harmonics in polarization (but not in strain) should vanish.[30] In experimental physics, perfect symmetry may be considered an idealization. Asymmetry may arise on different length scales of materials, reflecting physical processes or be induced artificially by constant external fields or other boundary conditions. The presence of a second harmonic component itself therefore may not serve as a robust indicator of a given mechanism or material's true symmetry, especially for a thin specimen were weak external forces can generate huge fields. However, variations and anomalies may yield valuable insights. An example of intentionally induced asymmetry is the alignment of ferroelectric domains by the application of electric fields larger than the coercive field, so called poling. Experimentally polarized samples show a second harmonic phase angle of polarization around +90° or -90° dependent on the orientation of the polarization with respect to the driving field.

The third harmonic phase angle, $\delta_3$, of polarization has been proven to be a sensitive indicator of different processes operating in ferroic materials. Phase angles between 0° and -240° (negative angles measured clockwise) have been observed experimentally and related to different underlying mechanisms. Polarization saturation in lead magnesium niobate and barium strontium titanate has been correlated to $\delta_3=0°$.[13,31] Phase angle of -90° has been correlated to displacement of ferroelectric and ferroelectric-ferroelastic domain walls in soft ferroelectric materials as described by Rayleigh-Néel model. [31,24,32] For materials with absence or a weak activity of extrinsic contributions such as hard ferroelectrics at weak fields, dielectric or piezoelectric materials like sapphire, strontium titanate at room temperature or quartz, or in many materials for weak driving fields, third harmonic phase angle around -180° is observed. Phase angle of -240° have been measured under high driving fields for hard-doped PZT were pinching of the polarization electric field loops becomes apparent[22] and for some lead-based relaxor-ferroelectric solid solutions when poled along the $[111]_{pc}$-direction. The influence of the third harmonic phase angle on the shape of the polarization electric field hysteresis loop for various third harmonic phase angle $\delta_3$ is illustrated in Figure 2.

An overview of commonly measured third harmonic phase angles and their evolution with increasing electric field amplitude are presented in the phasor diagram in the center of Figure 2. An example of a detailed analysis of the nonlinear polarization in (1-$x$)Pb(Mg$_{1/3}$Nb$_{2/3}$)O$_3$-$x$PbTiO$_3$ solid solution in terms of $\delta_3$ can be found in Ref.[7]. Evolution of $\delta_3(E)$ across the phase diagram together with microscopic scanning transmission electron microscopy studies aided in identification and discovery of a new microscopic mechanism contributing to the macroscopic properties of this technologically important material.

**V. THEORETICAL INTERPRETATION**

One example where there is a definite theoretical prediction of nonlinear behavior is the well-known Rayleigh law. The interaction of moving interfaces with pinning center is a general concept that has attracted broad interest in various fields of science. It was described empirically first for magnetic systems by Rayleigh[33], derived theoretically by Néel[23] and later by Kronmüller[34], introduced to ferroelectric systems by Turik[35], and discussed for the piezoelectric effect in ferroelectrics by Damjanovic[36]. Presently there is no firm experimental evidence that Rayleigh law is valid for strain-stress measurements in ferroelectric materials with ferroelectric-ferroelastic domains. In the only available study known to the authors, the Young modulus as a function of stress amplitude is not strictly a linear function[37], as required by the Rayleigh law. On the contrary, the available evidence (dependence of friction on stress amplitude) points that the law may not be valid for strain-stress relation in purely ferroelastic materials.[38] This raises interesting questions on the mechanisms behind the law's validity for ferroelastic-ferroelectric domain walls contributing to mixed electro-mechanical phenomena in ferroelectrics and its purported absence for purely ferroelastic walls in ferroelastics. The interaction of ferroelectric domain walls with pinning center was investigated from the early days of ferroelectricity. It's detailed understanding may allow for effective tailoring of properties.

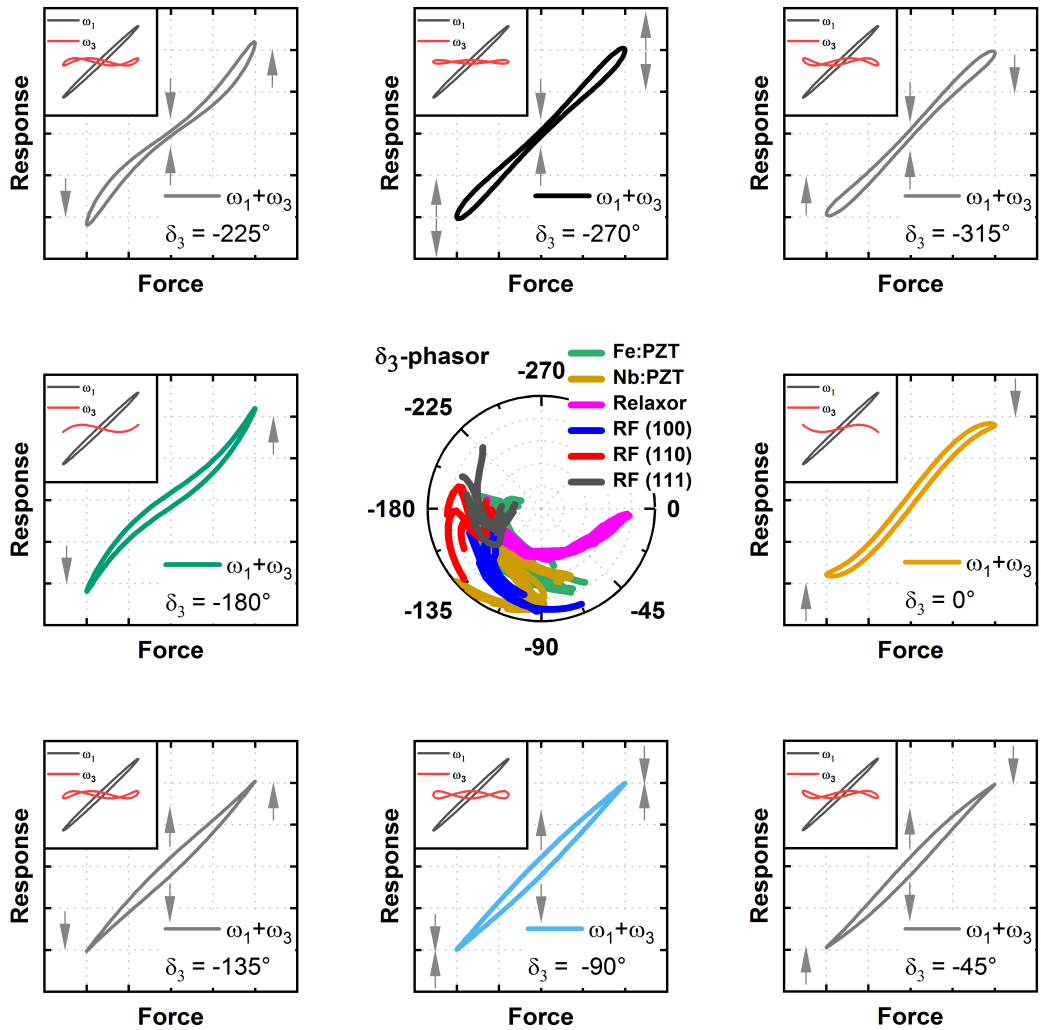

**Figure 2:** Geometric description of the combinations between first (arbitrarily taken as $\delta_1 = -5°$) and third harmonic components, and phasor diagram for commonly observed phase angles of the third harmonic of polarization $\delta_3$: 0°, indicating response saturation; -90°, indicating Rayleigh behavior; -180°, indicating polarization response of most nonferroelectric dielectrics and hard ferroelectrics; -270°, indicating polarization loop pinching and hard ferroelectrics (experimentally one usually observes -240° because of competing contribution from a dynamic component with $\delta_3=-180°$). The phasor diagram in the middle is constructed using more than 40 individual measurements as illustrated in Figure S2). RF stands for relaxor ferroelectrics: PZN-4.5PT, PZN-7PT, PMN-28PT, PMN-33PT. Relaxor: PMN ceramic, PMN crystal with [100]$_{pc}$ orientation. Nb:PZT stands for 1 at% Nb doped PZT ceramics with Zr/Ti ratio: 42/58, 52/48, 58/42. Fe:PZT stands for 1 at% Fe-doped PZT with Zr/Ti ratio: 42/58, 52/48, 58/42.

It was shown that domain wall movement in soft ferroelectrics results in polarization- and strain-electric field hysteresis loops that can be described by the Rayleigh equations[2,3,39]:

$$P(E) = (\varepsilon_{init} + \alpha_\varepsilon E_0)E \mp \frac{\alpha_\varepsilon}{2}(E_0^2 - E^2), \qquad (9)$$

and

$$x(E) = (d_{init} + \alpha_d E_0)E \mp \frac{\alpha_d}{2}(E_0^2 - E^2), \qquad (10)$$

where $E$ is an electric field value smaller or equal to the electric field amplitude $E_0$, $\varepsilon_{init}$ is the intrinsic relative dielectric permittivity, $\alpha_\varepsilon$ is the Rayleigh coefficient of dielectric permittivity, $d_{init}$ is the intrinsic piezoelectric coefficient, $\alpha_d$ is the Rayleigh coefficient of the piezoelectric coefficient. The "–" sign corresponds to the decreasing and "+" sign to the increasing field. Zero-field polarization and strain are given as $\mp \frac{\alpha}{2} E_0^2$, indicating hysteretic response. It is seen that the nonlinearity and hysteresis are directly coupled and completely described by the corresponding Rayleigh coefficient α. The Rayleigh law does not take into account rate-dependent loss mechanisms; the only contribution to hysteresis in equation 10 comes from irreversible jumps of domain walls.[11]

Equations 9 and 10 cannot *a priori* identify the underlying microscopic mechanisms. These may be revealed by the theoretical models[24,34] that derive these equations and that are subsequently confirmed in numerous, more or less direct experiments.[3,40] Some information about the dynamics of the unknown objects can be reached by considering purely formally predictions of the empirical equation along the lines discussed in Section IV Geometric interpretation. For example, the Fourier transform of equation 9, given in equation 11, indicates that the contributing objects may move reversibly (the linear component of the first harmonic) and irreversibly (all other terms) and that every irreversible, hysteretic motion is at the same time nonlinear (all terms containing α). Such insights pose a number of requirements on the model, as has been elegantly demonstrated by Néel in his derivation of the Rayleigh relations. Another, again purely formal insight, can be reached by taking Preisach approach[41,42,42,43,44] that identifies distribution of hypothetical internal and coercive fields associated with the units responsible for the Rayleigh behavior. In ferroelectric materials there is little doubt that the objects that dominate Rayleigh behavior are domain walls. The nature of the pinning centers that define the energy potential is less certain but essential for the understanding of microscopic mechanisms of nonlinearity and hysteresis. Kronmuller has related dislocation density in magnetic materials (pinning defects in that case) with the Rayleigh parameters[34] showing that model predicts α∼1/N, where N is the dislocation density. It is interesting to compare this prediction with experimental data in ferroelectric materials[45,15]. Hagemann reports data for acceptor doped $BaTiO_3$ for which Rayleigh law is not valid, but the slope of $\varepsilon(E)$ indeed decreases with increasing concentration of the dopant. Morozov *et al.* present data for both soft and hard PZT ($Pb(Zr_{1-x},Ti_x)O_3$), as shown in Figure S3. The soft PZT exhibits α∼N (N concentration of Nb) which seems to contradict the theoretical prediction. In fact, this may not be the case and the actual situation is more subtle. As the higher concentration of $Nb_{Ti}^{+1}$ most likely leads to a lower concentration of oxygen vacancies, $V_O^{+2}$, this behavior in fact may reveal that the actual pinning centers in both hard and soft PZT are oxygen vacancies. This illustrates how modeling of nonlinear properties in combination with experimental data can help identification of details of atomistic mechanisms operating in ferroelectrics.

The ideal Rayleigh law assumes motion of domain walls in a perfectly random energy potential which may be difficult to achieve in real systems. Thus, deviations are often observed, especially at very weak and strong fields. There are two approaches to explain this behavior. It is useful here to revoke results of the Preisach model. Put simply, it says that Rayleigh law arises in a system that consists of units which can be described by internal bias and coercive fields which have equal distribution (probability). So, a (small) departure from this ideal distribution would lead to a different dependence of the response on the field, but that does not mean that the mechanism is completely different than if this distribution were ideal.[46,47] In our opinion, it is justified to speak of Rayleigh-Néel like systems as long as hysteresis and nonlinearities are closely related, i.e., one can be calculated from the other.[39] If this is not the case, other processes in which nonlinearity and hysteresis emerge from different mechanisms may dominate.[15] It would be interesting to see which of these cases is valid for strain-stress relation in ferroelastic materials. The deviations from Rayleigh law may thus contain information on the distribution of pinning centers in a material, their strength, presence of non-domain related processes such as charge transport, and so on.[15]

To obtain criteria for the validity of Rayleigh parameters as a measure for extrinsic domain wall contributions it is convenient to convert it to an analytical form via expansion into Fourier series:

$$P(t) = (\varepsilon_{init} + \alpha_\varepsilon E_0) E_0 \sin(\omega t) + \sum_{1,3,5,\ldots} \frac{4\alpha_\varepsilon E_0^2 \sin\left(\frac{\pi n}{2}\right)}{\pi n(n^2-4)} \cos(n\omega t). \tag{11}$$

In this form several criteria for the justification of the use of Rayleigh equations for interpretation of data are given. The in-phase component of the permittivity and the piezoelectric coefficient of the first harmonic are linear functions of the driving field amplitude:

$$\varepsilon'^{(1)}(E_0) = (\varepsilon_{init} + \alpha_\varepsilon E_0), \varepsilon''^{(1)}(E_0) = -\frac{4\alpha_\varepsilon E_0}{3\pi}, \tag{12}$$

and

$$d'^{(1)}(E_0) = (d_{init} + \alpha_d E_0), d''^{(1)}(E_0) = -\frac{4\alpha_d E_0}{3\pi}. \tag{13}$$

All higher harmonic components are out-of-phase and quadratic functions of the driving field amplitude. In particular the phase angle of all higher harmonic is -90°. The amplitudes out-of-phase components decrease with increasing order of the harmonic and have specific ratios. For simplicity we shall refer to the real part of the first harmonic dielectric permittivity and the real part of the first harmonic piezoelectric coefficient as "dielectric permittivity" and "piezoelectric coefficient" and will denote them in the following as $\varepsilon^{(1)}$ and $d^{(1)}$ without prime symbols.

In practice the value of the third harmonic phase angle and the linear dependency of $\varepsilon^{(1)}(E_0)$ and $d^{(1)}(E_0)$ should be verified first. From Figure 2 it is seen that the condition of $\delta_3$ = -90° results in sharp tips of the corresponding hysteresis loop. Linear dependency of $\varepsilon^{(1)}(E_0)$ and $d^{(1)}(E_0)$ (nonlinear dependence of $P(E)$ and $x(E)$) can be evaluated by comparing the slope of subsequent loops with increasing driving field amplitude as shown for the dotted lines in Figure 1.

When the phase angle of the third harmonic is not -90° and the first-harmonic material coefficients are not linear functions of the field, this does not mean that the contributing process is not related to domain wall motion. It could indicate other mechanisms but also merely the fact that the domain walls move in a potential that is not random.[46] This is discussed in the next section.

**VI. PHENOMENOLOGICAL INTERPRETATION**

We illustrate this approach on representative ferroelectric, relaxor and relaxor ferroelectric materials. Lead zirconate titanate (PZT) is perhaps the most studied ferroelectric materials with arguably a rather good understanding of various contributions to polarization and strain[48,49,50,51]. Based on this knowledge and having in mind soft materials as a reference system for which theoretical description exists (i.e., Rayleigh law), we can attempt to interpret the cases where Rayleigh law does not hold. Pronounced nonlinearities are often related to reconfiguration of domain structures[22,52]. Domain wall movement can contribute to strain, polarization, or both depending on domain wall type and the symmetry of the property. Therefore dielectric and electro-mechanic nonlinearities can carry different information. Pb(Mg$_{1/3}$Nb$_{2/3}$)O$_3$ (PMN) is a canonical relaxor which has been widely modeled but, as we shall see, the models are incomplete as far as the nonlinear response is concerned. Finally, (1-$x$)Pb(Mg$_{1/3}$Nb$_{2/3}$)O$_3$-$x$PbTiO$_3$ (PMN-100$x$PT) and (1-$x$)Pb(Zn$_{1/3}$Nb$_{2/3}$)O$_3$-$x$PbTiO$_3$ (PZN-100$x$PT), are considered as examples of relaxor ferroelectrics.

VIa. Harmonic analysis of lead zirconate titanate

In Figure 3 the third harmonic phase angle of polarization of Nb- (soft) and Fe- (hard) doped rhombohedral lead zirconate titanate are compared for the dielectric permittivity and piezoelectric coefficient as a function of driving field amplitude.[15,50] Polarization and strain response were measured simultaneously as hysteretic nonlinear mechanisms may alter the state of the sample during measurements.[22] The inset in Figure 3a) shows the onset of nonlinear behavior for electric field strength of around 0.2 kV/cm as generally observed for soft doped PZT ceramics.[48,49,53] For soft and hard compositions $\delta_3$ of polarization is approximately -180° at weak fields as expected for weakly active extrinsic contributions. The $\delta_3$ of polarization of Nb-doped PZT approaches -90° (Rayleigh case) at electric field amplitudes of about 1 kV/cm before it is reduced to approximately -135° at around 5 kV/cm resulting in a broad peak around 1.5 kV/cm. Only marginal differences in the $\delta_3$ of polarization are seen for two soft-doped compositions. The field-dependent dielectric permittivity Figure 3b) and piezoelectric coefficient Figure 3c) show a linear increase up to approximately 1.5 kV/cm, which could be considered as an upper limit of Rayleigh like behavior. Note that linear polarization and strain response would be represented by constant permittivity and piezoelectric coefficient. For Fe-doped PZT, $\delta_3$ approaches -60° to -70°.

Comparing dielectric permittivity and piezoelectric coefficient of soft- and hard-doped PZT in Figure 3, larger extrinsic contributions for soft-doped samples become apparent. Moreover, it is seen that doping causes a larger variation in the absolute values of piezoelectric coefficients than in the dielectric permittivity. This indicates that doping affects extrinsic contributions that can have relatively stronger contributions to strain than to polarization, for example non-180° domain walls. Deviation from Rayleigh behavior and discrepancies between dielectric and piezoelectric non-linearities are clearly seen for rhombohedral soft-doped PZT around 2 kV/cm. Such anomalies were previously reported in ceramics[36] and films[54]. Studies of PZT films with different crystallographic orientations correlated deviations in dielectric and piezoelectric non-linearities to different contributions of ferroelectric and ferroelastic domains.[55] Rhombohedral (100)-oriented films, where 71° and 109° domains walls are not ferroelastically active, show stronger deviations between dielectric and piezoelectric non-linearities as compared to (111) oriented films, where 71° and 109° domains walls are ferroelastically active.

The deviation of permittivity and piezoelectric coefficient from linear field dependence (nonlinear strain and polarization) between 1 kV/cm and 2 kV/cm in Figure 3b) and c) could consequently be correlated to a pronounced onset of non-180° switching. These anomalies are more visible in the derivative of the dielectric permittivity shown in the inset of Figure 3b). A pronounced onset of non-180° switching may result in two competing effects: an increase in macroscopic polarization aligned with the field vector during each half cycle and a reduction of domain wall density. Considering the relatively small hysteresis between increasing and decreasing field subcycle in dielectric permittivity and piezoelectric coefficient, the change due to modifications of the macroscopic polarization state appears to be small. In the framework of Rayleigh equations the reduction of slope could be interpreted as the reduction of extrinsic contributions.[45] However, the onset of switching would effectively transfer response from the first to the second harmonic as domains can switch twice per bipolar cycle. A representation of the field dependence in terms of real and imaginary components of $P^{(3)}$ and $x^{(3)}$ could be helpful for additional analysis, as demonstrated in Ref.[7].

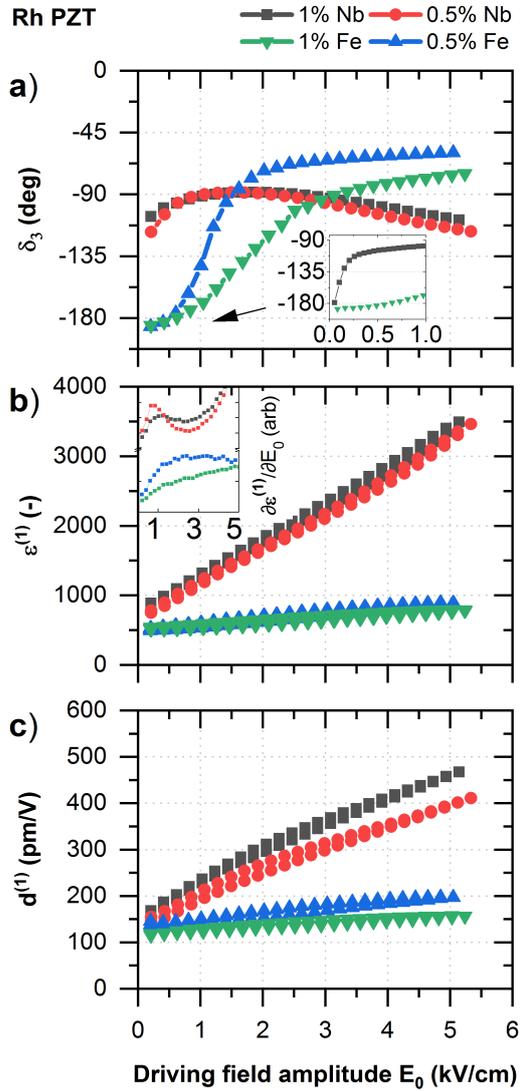

Figure 3: Dielectric and electro-mechanic nonlinearities of poled Nb- and Fe-doped rhombohedral lead zirconate titanate PZT(58/42). Dielectric and piezoelectric data were acquired simultaneously. a) Third harmonic phase angle of polarization $\delta_3$, b) dielectric permittivity, and c) piezoelectric coefficient. Each curve in b) and c) shows data for increasing and decreasing field half-cycle.

Normalized permittivity and piezoelectric coefficients presented in Figure 4 show that the relative deviation between nonlinear polarization and strain response increase towards rhombohedral compositions. These findings are consistent with the experimentally determined increase of extrinsic and non-180° switching contributions towards rhombohedral compositions.[56] Assuming the onset of pronounced non-180° domain wall switching as the origin for the observed anomaly and discrepancy between dielectric and piezoelectric nonlinearities, simultaneously occurring anomalies in the second harmonic components of polarization and strain can be expected. We note that the second harmonic is not expected from Rayleigh relations, however, in poled ceramics the symmetry is broken and the second harmonic may appear. It maybe manifested by different Rayleigh parameter $\alpha$ for the increasing and decreasing branches of the hysteresis loop (different irreversible behavior for increasing and decreasing field). In Figure 4b) the $\delta_2$ of polarization of PZT compositions doped with 1% Nb are presented. An anomaly in the phase angle of polarization is seen between 1 kV/cm and 2 kV/cm for the rhombohedral composition. For the tetragonal composition a similar anomaly in the phase angle of polarization could be measured between 2 kV/cm and 4 kV/cm. This shift could be explained with the higher coercive field and larger difference in spontaneous strain for non-180° switching of tetragonal compositions.

Consistent with the observed anomalies in the $\delta_2$ of polarization a marked increase in the amplitude of the second harmonic in polarization $P^{(2)}$ is observed at the corresponding fields as depicted in Fig. S4. For the composition in the vicinity of the morphotropic phase boundary (MPB) such clear anomalies could not be detected even though a subtle peak in the $\delta_2$ around 1 kV/cm can be revealed as shown in the inset of Figure 4a). It is assumed that a relatively stronger contribution of a different origin prevents more pronounced changes in $\delta_2$ of polarization. For a more detailed analysis it can be useful to analyze the real and the imaginary parts of harmonic components in addition, as done in Ref.[7].

Analyzing the real and imaginary parts of the second harmonic in polarization in more detail it can be shown that observed anomalies in $\delta_2$ of polarization are related to local maxima in the real part of the second harmonic component (Fig. S5). In contrast to rhombohedral and tetragonal samples the MPB composition shows a local minima followed by a strong increase. It appears that other contributions (e.g.

interphase motion and electrostriction) indeed dominate the second harmonic response. While the discussed observations are ascribed to domain wall movement, similar results might be obtained for mechanism of other origin.

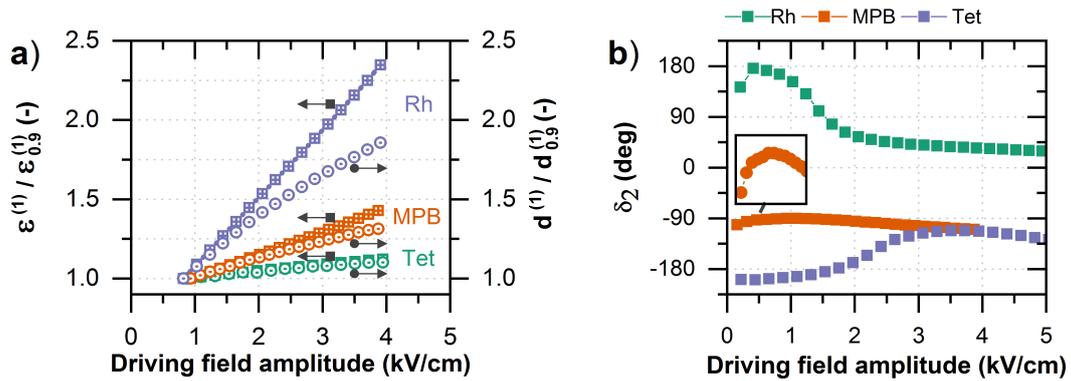

Figure 4: Dielectric and piezoelectric non-linearities for Nb:PZT. a) normalized dielecric permittivity and piezoelectric coefficient of Nb:PZT. The data are normalized to the measurement point at a driving field amplitude of ~0.9 kV/cm. b) Second harmonic phase angle of polarization for tetragonal (42/58) morphotropic (52/48), and rhombohedral (58/42) PZT ceramics with 1% Nb.

VIb. Harmonic analysis of lead magnesium niobate

Lead magnesium niobate (PMN) is considered the model system for relaxors and is still nurturing interest and dispute in the scientific community.[57,58,59,60,61,62,63,64] At ambient temperatures it features high values and distinct nonlinearities in the dielectric permittivity[31,65] in the absence of classic ferroelectric domains[66,67]. It was recently shown that PMN is weakly piezoelectric[58] in the unpoled state, ceramics can be poled to exhibit a rather large piezoelectric coefficient[68], is macroscopically polar at ambient temperature[6], and that a peak in nonlinear dielectric permittivity, as presented in Figure 5a) corresponds to macroscopic polarization reversal[6]. These findings were explained by the presence and the dynamic behavior of polar nano-entities (usually called polar nano regions, the objects whose properties are not well understood and for which there is no consensus in the literature).[6,58,61] The question therefore arises, how polarization reversal of polar nano-entities and ferroelectric domains compare and what can be learned from studies of nonlinearities.

Nonlinear dielectric properties of PMN ceramics are presented in Figure 5. In contrast to PZT a peak in dielectric permittivity around 1.4 kV/cm and a steep increase of $\delta_3$ from -180° to -0° can be observed. Taking into consideration the second harmonic response of PMN in Figure 5b), similarities between PMN and Nb-doped PZT ceramics (Figure S5) become evident. For both systems a local maximum in the real part of the second harmonic amplitude is correlated to an anomaly in the dielectric permittivity for the first harmonic. This anomaly is more visible in the derivative of the dielectric permittivity, as shown in the inset of Figure 3b). We propose therefore that a local maximum in the real part of the second harmonic component is common signature for polarization reversal of ferroic domains and polar nano nano-entities. This is not unexpected knowing that polarization reversal also contributes to the second harmonic of the strain.[69]

Given the general nature of polarization reversal additional information are required to analyze underlying mechanisms in more detail. Comparing $\delta_3$ of PMN in Figure 5a) with the examples given in Figure 2 saturation of polarization can be deduced ($\delta_3 = 0$). The fact that the permittivity drops after its maximum suggests in addition that the contribution of polar nano-entities at field amplitudes above 1.4 kV/cm to the permittivity has qualitatively changed. This is an interesting point that deserves additional discussion. A classical model of relaxors is superparaelectric model,[70] which describes reorientation of nanoregions in the dielectric field. An alternative model, so-called "breathing" model[71], is based on displacement of boundaries of the nanoregions in the electric field. The latter model makes a definite prediction about the phase angle of the third harmonic: it should be -180°, as observed experimentally.[71] However, this is valid only at very weak fields, Figure 5a. $\delta_3(E_0)$ approaching zero as the field amplitude increases indicates saturation of the contribution. This suggests that either displacement of the interfaces ("breathing") saturates at higher fields, or, it might mean that at higher fields polar nanoregions start flipping and it is this flipping mode that eventually saturates at high fields (once flipped the further increase of the field might only causes extension of the dipole). The latter possibility sounds reasonable knowing that the maximum in the permittivity is indeed correlated with the flipping of the macroscopic polarization of PMN. Thus, both breathing and flipping models may be correct, but at different field levels. Measurements of the corresponding electro-mechanic nonlinearities are expected to yield further insight and are currently undertaken.

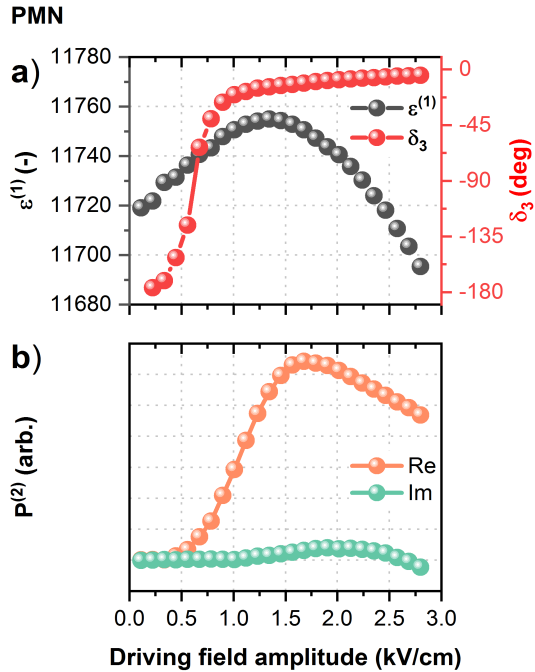

Figure 5: Dielectric nonlinearities of a Pb(Mg$_{1/3}$Nb$_{2/3}$)O$_3$ ceramic: a) First harmonic permittivity and third harmonic phase angle δ$_3$ of polarization. The peak in permittivity is attributed to reorientation of polar nano-entities.[6] b) Real and imaginary parts of second harmonic component of polarization. The local maximum in the real part is correlated to aforementioned reorientation of polar nano-entities.

VIc. Harmonic analysis of relaxor ferroelectric solid solutions

Relaxor ferroelectric solid solutions outperform classic ferroelectric and piezoelectric materials by far. Hierarchical polar structures and corresponding softening mechanism were proposed[7,72]. It was shown in the previous sections that distinct features in dielectric and electro-mechanic nonlinearities of ferroic domains and polar nano-entities (such as polar nanoregions in relaxors) exist. The question arises how domains and nanoentities interact if present in hierarchical polar structures. To avoid additional interactions with grain boundaries results of measurements discussed here were performed on singe crystals of different orientations.

Dielectric nonlinearities of various compositions and crystallographic directions of poled crystals of three different relaxor-ferroelectric solid solutions are presented in Figure 6. A strong anisotropy in the nonlinear response is apparent. For samples poled along the pseudo cubic $[001]_{pc}$-direction a linear increase in first harmonic permittivity and $\delta_3$ of polarization approaching -90° characteristic for Rayleigh behavior is observed. Contrasting these results, the nonlinear dielectric permittivity of samples poled along $[111]_{pc}$-direction is described by a higher order polynomial and a $\delta_3$ of polarization that features a peak around 0.4 kV/cm.

The relative slow increase of permittivity at low fields as compared to $[001]_{pc}$-poled samples and the drop of $\delta_3$ of polarization below -180° at higher fields indicate hardening and an unexpected tendency toward loop pinching in $[111]_{pc}$-poled samples (see Figure 2). Samples poled along the $[011]_{pc}$-direction show intermediate behavior. Qualitatively, a similar behavior is observed in all investigated compositions, although the nonlinearity of dielectric permittivity is weaker in the composition closer to the morphotropic phase boundary (PMN-33PT) as compared to the rhombohedral composition (PMN-28PT).

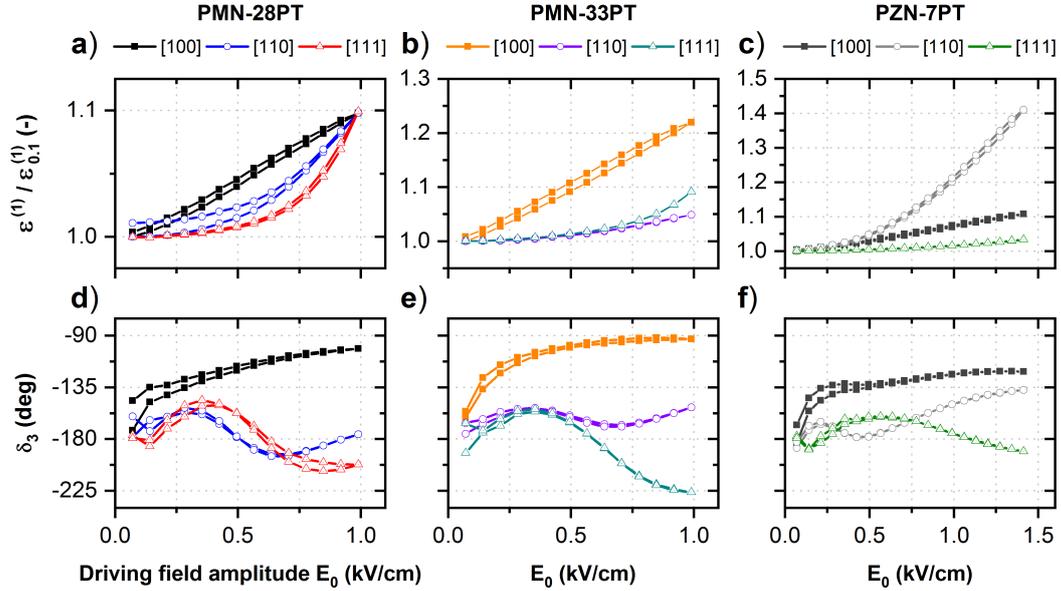

Figure 6: a)-c) Normalized dielectric permittivity (first harmonic) and d)-f) third harmonic phase angle $\delta_3$ as function of driving field amplitude for crystals poled along three pseudo cubic crystallographic directions. The permittivity data are normalized to the value measured at a driving field ampliude of ~0.1 kV/cm. Note that the ordinates of figures are not identical. a)-c) are not identice a&d) $Pb(Mg_{1/3}Nb_{2/3})O_3$–$0.28PbTiO_3$ (PMN-28PT), b&e) $Pb(Mg_{1/3}Nb_{2/3})O_3$–$0.32PbTiO_3$ (PMN-32PT), and c&f) $Pb(Zn_{1/3}Nb_{2/3})O_3$–$0.07PbTiO_3$ (PZN-7PT). $[100]_{pc}$-cut crystals approach Rayleigh behavior. $[111]_{pc}$-cut crystals show a peak in $\delta_3$ of polarization at relatively weak field levels before dropping below -180°. The latter is reminiscent of pinched hysteresis loops. $[110]_{pc}$-cut crystals show intermediate behavior.

To correlate the observed non-linearities with microscopic features, a $[111]_{pc}$-oriented $Pb(Mg_{1/3}Nb_{2/3})O_3$–$0.28PbTiO_3$ crystal (PMN–28PT) was polished to ~200 µm thickness and coated with transparent indium tin oxide electrodes. The sample was then electrically contacted to the top and bottom electrodes and mounted onto a stage of a polarized light microscope (PLM). Dielectric nonlinearities were first measured in the unpoled state before the picture presented in Figure 7a) was taken. The corresponding $\delta_3$ of polarization is shown as inset. A complex domain pattern and Rayleigh-like behavior is observed. The sample was then thermally depolarized and subsequently poled at room temperature. Dielectric nonlinearities of the poled state were measured and a PLM image (shown in Figure 7b) was taken again after the experiment. A clear contrast in the PLM image and a peak in the third harmonic phase angle (see inset) can be observed for the $[111]_{pc}$-poled PMN–28PT crystal, as previously described in Figure 6d).

Clear differences in the polar/domain structure are observed optically for unpoled and poled PMN-28PT $[111]_{pc}$-cut crystals after measurements of dielectric nonlinearities. The absence of contrast under the PLM for the poled case ~~shows~~ suggests that the mechanism responsible for the peak in $\delta_3$ of polarization (Figure 7b and 6d) acts on a shorter length scale as compared to the dominant mechanism in the unpoled state. Note that 180° domain walls cannot be seen in the case of PLM images but their density should be small in poled samples. However, considering the polar features that were recently visualized in PMN–$x$PT[73] it can be assumed that mechanisms on the nano-scale play a key role in the observed nonlinear dielectric behavior[6] of the investigated $[111]_{pc}$-poled crystals. In addition the hardening effect of poling can be confirmed. While complex domain structures are present after field cycling in the case of an unpoled sample, the poled sample appears to be less affected, consistent with the drop in $\delta_3$ of polarization below -180° above 0.4 kV/cm.

Assuming polar nano-entities as the origin of the peak in the third harmonic phase angle, the second harmonic of polarization and strain are analyzed with respect to the previously discussed switching signatures. The real part of the second harmonic in strain and polarization are compared in Figure 8 for $[111]_{pc}$- and $[011]_{pc}$-poled lead zinc niobate lead titanate crystals (PZN–7PT). Both samples show a local extreme in the real part of the second harmonic polarization as previously observed for PZT and PMN, indicating a switching mechanism of active substructures (polar nano-entities) within ferroic domains. In addition, a local extreme in the real part of the second harmonic in strain is present as well.

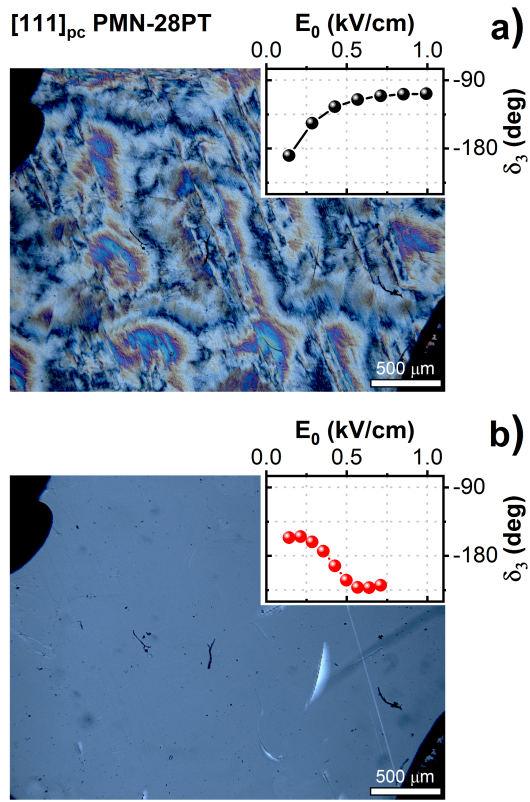

**Figure 7:** Polarized light microscopy (PLM) images and third harmonic phase angle $\delta_3$ of polarization of a ~200 μm thick $[111]_{pc}$-cut $Pb(Mg_{1/3}Nb_{2/3})O_3$–$0.28PbTiO_3$ (PMN-28PT) crystal with transparent indium tin oxide electrodes. The PLM images were taken after the measurements of $\delta_3$ shown in the insets. The sample was thermally depolarized between measurements: a) initally unpoled sample shows Rayleigh like $\delta_3$ behavior and complex domain pattern after the measurement, b) the same sample as in a) but poled before the measurements of $\delta_3$. The absence of clear polar features suggests that the mechanism responsible for the peak in $\delta_3$ has dimensions below the resolution limit of visible light.

Similar results for polarization and strain response provide strong evidence that the underlying mechanism is electro-mechanically active. However, even if the results presented suggest that polar nano-entities switch within parent domains (based on anomalies in the second harmonic, Figure 8), the absence of obvious anomalies in the dielectric permittivity (Figure 6) at all electric field levels investigated suggests that the direct contribution of such a dynamic mechanism is relatively weak and that it is rather the presence of polar nano-entities that affects the dielectric response by softening the material.[72,74] This is somewhat analogous to the case where domain walls improve properties of a ferroelectric material not by their displacement but by their presence. This has been observed, for example, in $BaTiO_3$ single crystals where stationary charged domain walls improve piezoelectric and dielectric response by destabilizing polarization within domains even if they do not move.[75,76])

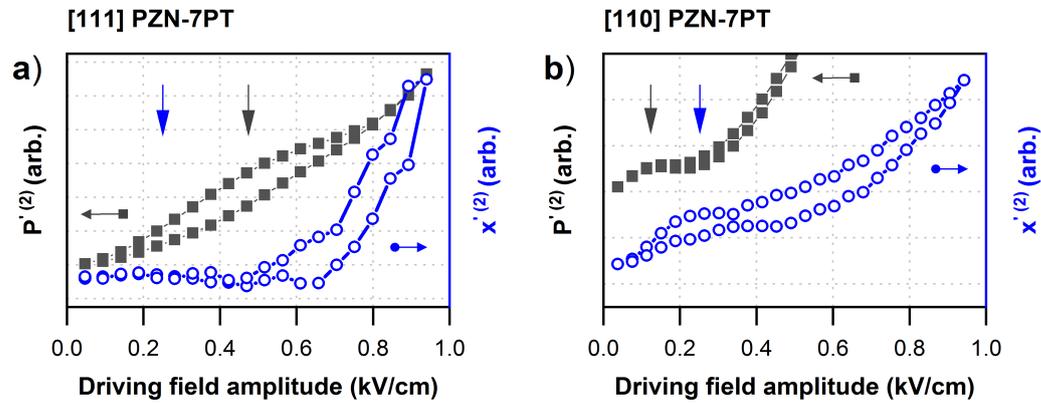

Figure 8: Real part of the second harmonic of polarization and strain for Pb(Zn$_{1/3}$Nb$_{2/3}$)O$_3$–0.07PbTiO$_3$ (PZN-7PT) crystals: a) Poled along the [111]$_{pc}$ and b) [011]$_{pc}$ crystallographic direction. Anomalies in the polarization and strain response are indicated as vertical black and blue arrows, respectively. They are interpreted as switching of electro-mechanically active substructures within ferroic domains.

## VII. CONCLUDING REMARKS

Dielectric and electro-mechanic nonlinearities are of key importance for understanding the functioning of materials' used for modern electronic devices. The interpretation and prediction of material performance is in general difficult. This article presents an approach to correlate signatures in the dielectric and the electro-mechanic nonlinearities to the underlying physical mechanisms. Fundamental formalism and concepts are introduced and demonstrated on model ferroelectric, relaxor and relaxor ferroelectric systems.

**SUPPLEMENTARY MATERIAL**
Additional comments on formal description, the second harmonic polarization response and polarization in relaxor PMN are available in the supplementary material.

ACKNOWLEDGMENTS
This work was supported by the Swiss National Science Foundation (No. 200021_172525).
We would like to thank Dr. Nigon Robin for the deposition of indium thin oxide electrodes and TRS technology for supplying some of single crystals used in the study.

AIP PUBLISHING DATA SHARING POLICY

The data that support the findings of this study are available within the article and its supplementary material. Raw data of this study are available from the corresponding author upon reasonable request.

Supplementary Material: Dielectric and electro-mechanic nonlinearities in perovskite oxide ferroelectrics, relaxor and relaxor ferroelectrics

Lukas M. Riemer[1,a)], Li Jin[2], Hana Uršič[3], Mojca Otonicar[3], Tadej Rojac[3], and Dragan Damjanovic[1].

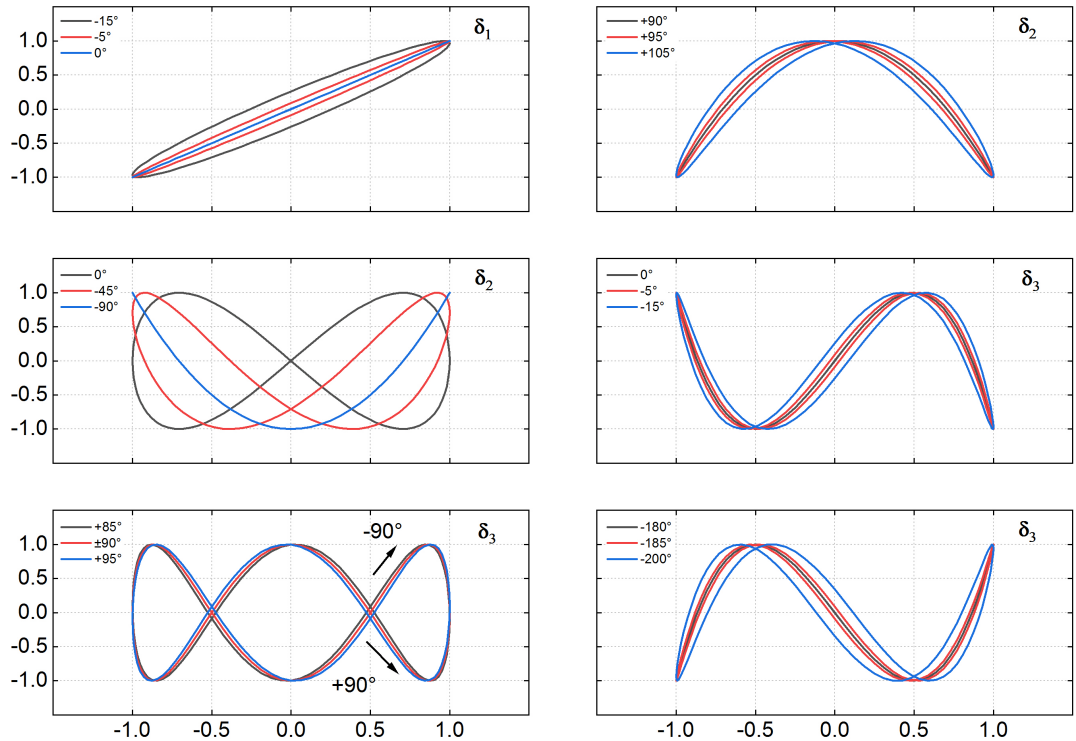

**Figure S1:** The first three harmonic components of P(E) at different values of the phase angle $\delta_n$. The vertical axis is $P_n \sin(n\omega t + \sigma_n)$ and the horizontal $E_0 \sin(\omega t)$.

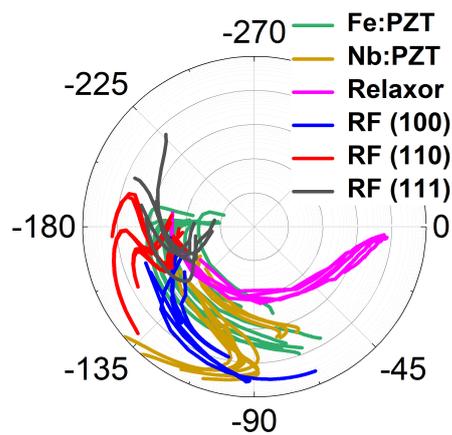

**Figure S2:** Phasor diagram for commonly observed phase angles of the third harmonic of polarization. $\delta_3$: 0°, indicating response saturation; -90°, indicating Rayleigh behavior; -180°, indicating polarization response of most nonferroelectric dielectrics, hard ferroelectrics, relaxors and many ferroelectrics at very weak fields; -270°, indicating polarization loop pinching and hard ferroelectrics (experimentally

one usually observes -240° because of competing contribution from a dynamic component with $\delta_3$=-180°). RF stands for relaxor ferroelectrics: PZN-4.5PT, PZN-7PT, PMN-28PT, PMN-33PT. Relaxor: PMN ceramic, PMN crystal with $[100]_{pc}$ orientation. Nb:PZT stands for 1 at% Nb doped PZT ceramics with Zr/Ti ratio: 42/58, 52/48, 58/42. Fe:PZT stands for 1 at% Fe-doped PZT with Zr/Ti ratio: 42/58, 52/48, 58/42.

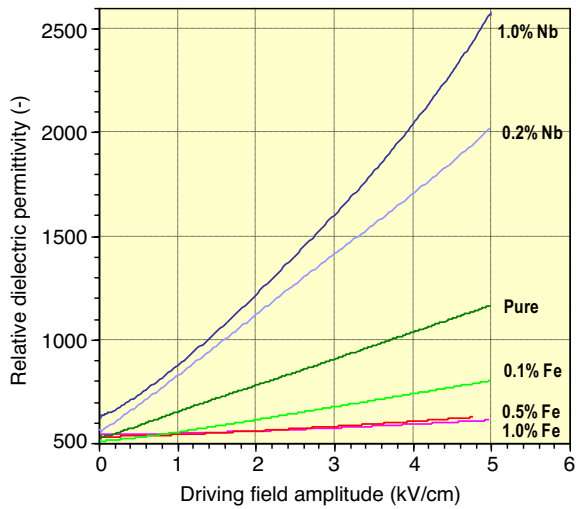

Figure S3: Relative dielectric permittivity as a function of the driving field amplitude for soft (Nb-doped), "pure" (undoped) and hard (Fe-doped) $Pb(Zr_{0.58}Ti_{0.42})O_3$ ceramics. After Morozov.[1]

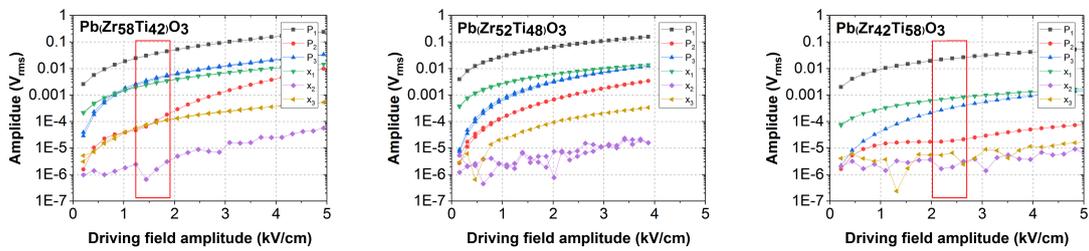

Figure S4: Amplitudes of the first three harmonic components in strain and polarization for different PZT compositions with 1% Nb-doping. The $V_{rms}$ values on the ordinate axis indicates output of the lock-in amplifier, used to measure strain and polarization. The red rectangle marks the field region where the amplitude of the second harmonic in polarization exhibits a sharp increase.

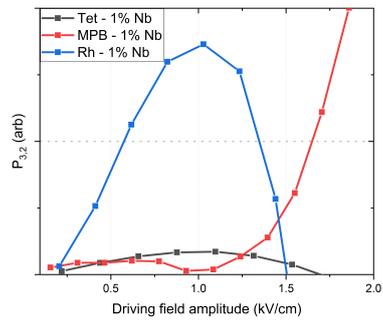

**Figure S5: Anomaly in the amplitude of the real part of the second harmonic in polarization of the Nb-doped PZT compositions.**

References
[1] M. Morozov, D. Damjanovic, and N. Setter, Journal of the European Ceramic Society **25**, 2483 (2005).